\documentclass[aps,amsmath,amssymb,amsfonts,superscriptaddress,floatfix]{revtex4} 
\usepackage{amsmath}
\usepackage{graphicx}
\usepackage{amsthm}
\newtheorem{theorem}{Theorem}
\newtheorem{lemma}{Lemma}
\newtheorem{corollary}{Corollary}
\newcommand{\sgn}{\sigma}
\newcommand{\He}{H^{empty}}

\newtheorem{definition}{Definition}
\newcommand{\be}{\begin{equation}}
\newcommand{\ee}{\end{equation}}

\newcommand{\calS}{{\cal S }}

\newcommand{\calD}{{\cal D }}

\newcommand{\ZZ}{{\mathbb{Z}}}

\begin{document}
\title{The Stability of Free Fermi Hamiltonians}
\author{M.~B.~Hastings}
\affiliation{Station Q, Microsoft Research, Santa Barbara, CA 93106-6105, USA}
\affiliation{Quantum Architectures and Computation Group, Microsoft Research, Redmond, WA 98052, USA}

\begin{abstract}
Recent results have shown the stability of frustration-free Hamiltonians to weak local perturbations, assuming several conditions.  In this paper, we prove the stability of free fermion Hamiltonians which are gapped and local.
These free fermion Hamiltonians are not necessarily frustration-free, but we are able to adapt previous work to prove stability.  The key idea is to add an additional copy of the system to cancel
topological obstructions.  We comment on applications to quantization of Hall conductance in such systems.
\end{abstract}
\maketitle

\section{Background}
\label{bkgnd}
Consider a local many-body Hamiltonian $H$ with a spectral gap.  We may define a ``phase of matter" to consist of all Hamiltonians $H'$ which are connected to $H$ by a path of local, gapped Hamiltonians\cite{hastingswen}.  In this case, quasi-adiabatic continuation\cite{lsmh} allows one to control many properties of $H'$; in fact, quasi-adiabatic continuation yields quantitative bounds which depend on the magnitude of the gap and on the interaction strength and range along the path, so that the bounds can be applied to finite systems, allowing us to sidestep the need to define families of Hamiltonians or consider systems of infinite size as would otherwise be needed to define a phase.

However, how can we prove that $H'$ is connected to $H$ by a gapped path?  One important case is when $H'$ is equal to $H$ plus some small local perturbation.  If the spectral gap of $H$ (difference between ground and first excited state) is equal to $\Delta$, then trivially if the norm of the perturbation is smaller than $\Delta$, the gap remains open along a linear path from $H$ to $H'$.  However, the perturbation will typically be a sum of local terms, each with strength bounded by some $J$.  If the number of local terms is proportional to system size, $N$, then this trivial bound is only useful for $J \lesssim \Delta/N$ and so is not useful for large $N$.  Instead, we seek bounds uniform in system size.

Such bounds have been proven for various systems which are sums of commuting projectors\cite{bhm,bh} and for frustration-free systems\cite{mff}.  However, these bounds have not yet been proven for one of the most important systems in physics, a system of gapped, local fermions.  
In this paper, we show a stability theorem for this system; the key step is to introduce an extra copy of the system which allows us to cancel certain topological obstructions.

We consider a free fermion Hamiltonian $H_{FF}$.  We use the most general form for such a Hamiltonian, writing it in terms of Majorana operators:
\be
\label{HFFdef}
H_{FF}=\sum_{j,k\in \Lambda} \gamma_j A_{j,k} \gamma_k,
\ee
where $A$ is an $N$-by-$N$ Hermitian antisymmetric matrix (hence, $A$ is pure imaginary), and $\gamma_j$ are Majorana operators obeying the relations $\{\gamma_j,\gamma_k\}=2\delta_{j,k}$, and $A_{j,k}$ denotes the $j,k$ matrix element of $A$. 
The labels $j,k$ label different sites arranged on a cubic lattice $\Lambda$ in $D$ dimensions.  We assume an exponential decay of matrix elements of $A$, defined more precisely below.

Let the smallest eigenvalue of $A$ in absolute value be equal to $\Delta/4$.  Then, $H_{FF}$ has a spectral gap between ground state and first excited state equal to $\Delta$ (the factor of $4$ follows after some algebra; the algebra may be more clear after Eq.~(\ref{Hedoub}) later).
 We take $N$ even (otherwise, $\Delta=0$), so the Fock space has dimension $2^{N/2}$.

Before giving our construction, we motivate it by considering some other possible attempts.
To describe these attempts, let us for simplicity consider
instead a Hamiltonian without superconducting pairing terms: $H_{FF}=\sum_{j,k\in \Lambda} \psi^\dagger_j T_{j,k} \psi_k$, where $\psi^\dagger,\psi$ are fermionic creation and annihilation operators.
Note that one can trivially bring $H_{FF}$ into a frustration-free form, at the cost of breaking locality badly, by finding a unitary transformation that diagonalizes $T$. Then, $H_{FF}=\sum_\alpha \epsilon_\alpha a^\dagger_\alpha a\alpha$,
where $a^\dagger_\alpha,a_\alpha$ label the transformed creation and annihilation operators and $\epsilon_\alpha$ are eigenvalues of $T$.
The ground state minimizes every term separately, but the orthogonal transformation may be highly nonlocal so that the operators $a_\alpha$ are nonlocal combinations of the operators $\psi_j$ so that a local perturbation may appear nonlocal in this basis.

One could instead hope to find an orthogonal basis of free fermion states which are
approximately local and which span the space of empty states.
If one could find such a basis, then one could make a frustration-free and local form of the ``spectrally flattened Hamiltonian"\cite{kitaevkthy},
where the spectrally flattened Hamiltonian has all positive eigenvalues replaced by $+1$ and all negative eigenvalues replaced by $-1$.  Such a Hamiltonian would be   $H_{FF}=\sum_\alpha a^\dagger_\alpha a_\alpha$, where $a_\alpha$
are annihilation operators for states in this basis.
However, such an orthogonal basis does not exist for Hamiltonians with certain topological obstructions although it does exist in the absence of these obstructions\cite{kitaevkthy,torus}.

It is possible that instead there exists an overcomplete and non-orthogonal basis of states for the empty band.
To apply theorems on frustration-free systems, however, we want this basis to be strictly local: every state in the basis should be supported on some region whose diameter is bounded in a way independent of system size.
This is a very strict requirement and it is not clear whether or not such a basis exists.  It is related to the question of
whether or not topologically nontrivial systems can have strictly local Green's functions, a question which seems to have a negative answer in some cases.  One may be able to modify some of the results to handle approximately local basis states.
Even if we could handle all these problems, this still only solves the case of finding a frustration-free form for the spectrally flattened Hamiltonian, rather than the original Hamiltonian.
(We remark that while this paper was in preparation, another paper\cite{NSY} appeared that previewed a stability proof based on ideas similar to those suggested in the above paragraphs.  However, it appears that the authors of that paper note similar problems that will arise and do not have a means to avoid all these problems, which, as far as I understand their work, means that the results that they preview will be restricted in the possible systems that can be considered.)

Given these problems,we use a different approach.
Our basic idea is to add
another copy of the system to cancel topological obstructions.  More precisely, we define a Hamiltonian $H_{doub}$ which consists of two copies of $H_{FF}$ with {\it opposite signs}:
\be
\label{Hddef}
H_{doub}=\sum_{j,k\in \Lambda} \gamma_{j,1} A_{j,k} \gamma_{k,1} -\sum_{j,k\in \Lambda} \gamma_{j,2} A_{j,k} \gamma_{k,2},
\ee
and we show stability of $H_{doub}$ which implies stability of $H_{FF}$.
Now, the Majorana operators $\gamma_{j,a}$ are indexed by a pair $j,a$, where $j$ labels a site and $a\in\{1,2\}$.

The Hamiltonian $H_{doub}$ is topologically trivial; indeed, following Ref.~\onlinecite{z2doub}  one can
give an explicit path of free fermion Hamiltonians connecting $H_{doub}$ to a gapped, diagonal free fermion Hamiltonian.  However, we will not use this path in the present paper.
Instead we give a unitary map which approximately preserves locality and maps $H_{doub}$ to a Hamiltonian $\He$.  The Hamiltonian $\He$ physically describes a system without superconducting pairing whose ground state is the empty band.
This unitary mapping maps an approximately local perturbation of $H_{doub}$ to an approximately local perturbation of $\He$.
We then show stability of $\He$ to that perturbation, by modifying the proof of \cite{bh}; that paper considers Hamiltonians which are sums of commuting projectors rather than merely frustration-free, but it will be possible to adopt it for our case by writing the Hamiltonian $\He$ as a sum of commuting terms (each term proportional to a number operator) plus a non-negative term.  We comment in section \ref{remarks} on why we choose to show stability of $\He$ this way.

\section{Definitions and Results}
We consider systems defined on a $D$-dimensional cubic lattice $\Lambda$.
The system will have a finite linear size $L$, but all bounds that we prove will be uniform in size.
Each site will be labeled by an index $j$; when we consider $H_{FF}$, there will be one Majorana operator $\gamma_j$ for each site $j$
and  and when we consider $H_{doub}$ there will be two
Majorana operators $\gamma_{j,1}$ and $\gamma_{j,2}$ for each site $j$.  
The case of a Hamiltonian with many Majorana operators per site can be handled as follows: replace a single site with many modes by many sites, each with one mode, and appropriately rescale the decay constants $\mu,\nu$ below.  We discuss this further in section section \ref{remarks}, considering uniformity with respect to the number of modes.
In some intermediate results below, we will allow an arbitrary number of Majorana operators per site for generality.

We consider a Hamiltonian
\be
H=H_{FF}+V,
\ee
where $H_{FF}$ is as defined in Eq.~(\ref{HFFdef}), with $H_{FF}$ having smallest eigenvalue in absolute value which is $\geq \Delta/4$.

We now define certain locality properties.  We use the term ``operator" for an operator acting on Fock space and we use ``matrix" for a matrix with one entry per Majorana operator (an $N$-by-$N$ matrix for $H_{FF}$ and a $2N$-by-$2N$ matrix for $H_{doub}$).
We express locality properties of Hamiltonians via a decomposition as a sum of terms;
 we shall often identify a Hamiltonian and the corresponding decomposition
unless it may lead to confusion.
\begin{definition}
We say that an operator $W$ has $(J,\mu)$ decay if $W$ can be decomposed as:
\be
\label{Wdecay}
W=\sum_{r \geq 1} \sum_{C \in {\cal S}(r)} W_{r,C},
\ee
where
${\cal S}(r)$ denotes the set of cubes of linear size $r$ and $W_{r,C}$ is an operator supported on a cube $C$
and where
\be
{\rm max}_{C \in {\cal S}(r)} \Vert W_{r,C} \Vert \leq J \exp(-\mu r),
\ee
for some constants $J,\mu>0$.
\end{definition}

\begin{definition}
With a single Majorana operator per site, we say that an $N$-by-$N$ matrix $B$ has $[K,\nu]$ decay if
\be
\label{Adecay}
|B_{j,k}| \leq K \exp(-\nu {\rm dist}(j,k)),
\ee
for some constants $K,\nu>0$, where ${\rm dist}(.,.)$ denotes the distance with the Manhattan metric.
In general, we say that a matrix $B$ has $[K,\nu]$ decay if, given any pair of sites $j,k$, the submatrix of $B$ whose row entries correspond to operators on site $j$ and whose column entries correspond to site $k$ has an operator
norm bounded by $K \exp(-\nu {\rm dist}(j,k))$,
for some constants $K,\nu>0$.
\end{definition}

These conditions are sufficient to imply a Lieb-Robinsion bound for $H_{doub}+V$, with commutators replaced by anti-commutators appropriately (consider only operators which are ``bosonic", i.e., a sum of products of an even number of Majorana operators, or ``fermionic, i.e., a sum of products of an odd number of Majorana operators; then, one uses anti-commutators when considering a pair of fermionic operators, and uses commutators otherwise).
The generalization from commutators to anti-commutators is standard in the theory of Lieb-Robinson bounds.  See for example Refs.~\onlinecite{ac1,ac2}.  Note that if we considered a system with an arbitrary number of Majorana operators associated with each site, then given just an assumption that $A$ has $[K,\nu]$ decay and $V$ has $(J,\mu)$ decay a Lieb-Robinson bound does not follow using known results; see remarks in section \ref{remarks}.  

All Hamiltonians, and all terms in the decomposition of a Hamiltonian, will be bosonic.

Our main result is that:
\begin{theorem}
\label{main}
There exist constants $J_0,c_1>0$ depending only on $K,\nu,J,\mu,\Delta,D$ such that
for all $J \leq J_0$, if $A$ has $[K,\nu]$ decay and $V$ has $(J,\mu)$ decay, then the Hamiltonian $H_{doub}+V$ has a spectral
gap between ground and first excited states which is greater than or equal to $\Delta-c_1 J$.
\end{theorem}

Theorem \ref{main} has the immediate corollary:
\begin{corollary}
There exist constants $J_0,c_1>0$ depending only on $K,\nu,J,\mu,\Delta,D$ such that
for all $J \leq J_0$,  if $A$ has $[K,\nu]$ decay and $V$ has $(J,\mu)$ decay, then  the Hamiltonian $H_{FF}+V$ has a spectral
gap between ground and first excited states which is greater than or equal to $\Delta-c_1 J$.
\begin{proof}
Take two copies of the system with Hamiltonian $H_{FF}+V$ acting on the first copy and Hamiltonian $-H_{FF}$ acting on the second copy.
Apply theorem \ref{main} to this system with two copies to get a lower bound on the spectral gap.  Since the two copies do not interact with each other, the spectrum of the system with two copies is the convolution of the spectra of each copy, and so a gap for the  system with two copies implies a gap for the first copy.
\end{proof}
\end{corollary}

To prove theorem \ref{main}, we will reduce the problem to a special case, theorem \ref{emptystable} below.  Physically, this special case will correspond to a free fermion Hamiltonian with no superconducting pairing terms and such that the ground state is the empty band.  The reduction to this special case is in section \ref{reduce}.

We define creation and annihilation operators $\psi^\dagger_j,\psi_j$ to be a set of operators that obey the anti-commutation relations $\{\psi^\dagger_j,\psi^\dagger_k\}=\{\psi_j,\psi_k\}=0$ and $\{\psi^\dagger_j,\psi_k\}=\delta_{j,k}$.  If $N$ is even, a system with $N$ Majorana operators can be re-expressed in terms of $N/2$ creation and $N/2$ annihilation operators by forming appropriate linear combinations; see Eq.~(\ref{Diracdef}) below.  We consider a system where for each site $j$ there is one pair of creation and annihilation operators $\psi^\dagger_j,\psi_j$.

We will
consider Hamiltonians
\be
\He=\sum_{j,k\in\Lambda} \psi^\dagger_j T_{j,k} \psi_k,
\ee
where $T$ is a Hermitian positive semi-definite matrix, with the smallest eigenvalue of $T$ lower bounded by $\Delta$.
We call this Hamiltonian $\He$ because its ground state is the ``empty state": $\psi_j$ annihilates the ground state for all $j$.

Then,
\begin{theorem}
\label{emptystable}
There exist constants $J_0,c_1>0$ depending only on $K,\nu,J,\mu,\Delta,D$ such that
for all $J \leq J_0$, if $T$ has $[K,\nu]$ decay and $V$ has $(J,\mu)$ decay, the Hamiltonian $\He+V$ has a spectral
gap between ground and first excited states which is
greater than or equal to $\Delta-c_1 J$.
\end{theorem}
We prove theorem \ref{emptystable} in section \ref{stabHempty} by modifying techniques from \cite{bh}.

\section{Doubled Hamiltonian}
\label{reduce}
In this section, we show that theorem \ref{main} follows from theorem \ref{emptystable}.
This is done by explicitly constructing a unitary transformation from a Hamiltonian obeying the conditions of theorem \ref{main} to one obeying the conditions of theorem \ref{emptystable}.

For any Hermitian matrix $B$ with no zero eigenvalues, let $\sgn(B)$ be the sign function applied to $B$: $\sgn(B)$ is $+1$ on the positive eigenspace of $B$ and $-1$ on the negative eigenspace of $B$.  If $B$ has eigenvectors $v_i$ with eigenvalues $\lambda_i$, let $|B|$ have eigenvectors $v_i$ with eigenvalues $|\lambda_i|$.
Note that $\sgn(A)$ is anti-symmetric and $|A|$ is symmetric.

Let
\begin{eqnarray}
O_{FF}=\frac{1}{\sqrt{2}}\begin{pmatrix}I &i\sgn(A) \\
i\sgn(A) & I \end{pmatrix}.
\end{eqnarray}
Since $\sgn(A)$ is anti-symmetric, $O_{FF}$ is orthogonal (note that since $\sgn(A)$ is pure imaginary, $i\sgn(A)$ is real).
Then,
\begin{eqnarray}
\label{singlep}
O^\dagger_{FF} \begin{pmatrix} A & \\ & -A \end{pmatrix} O_{FF} = \begin{pmatrix} 0 & i|A| \\ -i|A| & 0 \end{pmatrix}.
\end{eqnarray}

Eq.~(\ref{singlep}) is an equation involving $N$-by-$N$ matrices.  There is an analogous equation acting on the Fock space.  Since $O_{FF}$ is orthogonal, there is a unitary $U$ acting in the Fock space 
so
that
\begin{eqnarray}
\label{xformm}
U_{Fock}^\dagger \gamma_{j,1} U_{Fock} = \frac{1}{\sqrt{2}} \Bigl( \gamma_{j,1}+i \sum_{k \in \Lambda} (\sgn(A))_{j,k} \gamma_{k,2}\Bigr),
\\ \nonumber
U_{Fock}^\dagger \gamma_{j,2} U_{Fock} = \frac{1}{\sqrt{2}} \Bigl( \gamma_{j,2}+i \sum_{k \in \Lambda} (\sgn(A))_{j,k} \gamma_{k,1}\Bigr),
\end{eqnarray}
where $(\sgn(A))_{j,k}$ denotes the $j,k$ matrix element of $\sgn(A)$.
Hence,
\be
\label{Hdsub}
U_{Fock}^\dagger H_{doub} U_{Fock} =\He,
\ee
where
\be
\He=2i \sum_{j,k\in \Lambda} \gamma_{j,1} (|A|)_{j,k} \gamma_{k,2}.
\ee

Now define operators
\begin{eqnarray}
\label{Diracdef}
\psi^\dagger_j &=& \frac{\gamma_{j,1}-i \gamma_{j,2}}{2}, \\
\psi_j&=& \frac{\gamma_{j,1}+i\gamma_{j,2}}{2}.
\end{eqnarray}
Then,
\be
\label{Hedoub}
\He=4\sum_{j,k\in \Lambda} \psi^\dagger_j (|A|)_{j,k} \psi_k-{\rm const.},
\ee
where the constant is equal to $2{\rm tr}(|A|)$.

We claim that the matrix $\sgn(A)$
has $[M,\tau]$ decay
for some constants $M,\tau>0$ that depend only on $K,\nu,\Delta,D$.
Indeed, this decay property for $\sgn(A)$ follows from the exponential decay of correlations in a gapped system\cite{lsmh} obeying a Lieb-Robinson bound, since $(\sgn(A))_{j,k}$ is proportional to the expectation value of the commutator $[\gamma_j, \gamma_k]$ in the ground state of $H_{doub}$.  The matrix $\sigma(A)$ is also known as the spectrally-flattened Hamiltonian; see also Ref.~\onlinecite{kitaevkthy}. 

Given this decay of $\sgn(A)$, we claim that $4(|A|)_{j,k}$ also obeys an exponential decay bound so that setting $T=4|A|$, then $T$ has $[K',\nu']$ decay, where the constants $K',\nu'>0$ depend only on
$K,\nu,\Delta,D$.  To show this, note that
\begin{eqnarray}
|(|A|_{j,k})| &=& |\sum_{l\in\Lambda} \sgn(A)_{j,l} A_{l,k}| \\ \nonumber
&\leq & \sum_{l\in \Lambda} |\sgn(A)_{j,l}| |A_{l,k}| \\ \nonumber
&\leq & M K \sum_{l\in\Lambda} \exp(-\tau {\rm dist}(j,l)) \exp(-\nu {\rm dist}(l,k)).
\end{eqnarray}
Summing over $l$ in a cubic lattice gives an exponential decay in ${\rm dist}(j,k)$ for appropriate $K',\nu'$.
(One can also prove the decay of $|A|_{j,k}$ more directly using Fourier transforms, smooth approximations to the absolute value function, and Lieb-Robinson bounds; we omit this here.)

Also, in lemma \ref{Vprimedecay} below we show that if we set $V'=U_{Fock}^\dagger V U_{Fock}$, then $V'$ has $(J',\mu')$ decay, where the constants $J',\mu'>0$ depend only on $K,\nu,\Delta,J,\mu,D$.
Hence, $U_{Fock}^\dagger (H_{doub}+V) U_{Fock}$ obeys the conditions of theorem \ref{emptystable}, reducing the problem of proving theorem \ref{main} to that of proving theorem \ref{emptystable}.

For the next lemma, we allow an arbitrary number of Majorana operators per site for generality.
\begin{lemma}
\label{Vprimedecay}
Let there be $n_{max}$ Majorana operators per site.
Let $W$ have $(J_W,\mu_W)$ decay.  Let $U$ be a unitary with the property that
\be
\label{uupdate}
U^\dagger \gamma_{j,a} U=\sum_{k\in \Lambda} \sum_{b\in\{1,\ldots,n_{max}\}} O_{(j,a),(k,b)} \gamma_{k,b},
\ee
for some orthogonal $O$ with $[K_0,\nu_0]$ decay, where $1 \leq b \leq n_{max}$. Let $W'=U^\dagger W U$.  Then, $W'$ has $(J',\mu')$ decay, where the the constants $J',\mu'>0$ depend only on $J_W,\mu_W,K_0,\nu_0,n_{max},D$.

\begin{proof}
Consider any operator $W_C$ supported on some cube $C$.  Let $b_l(C)$ for any cube $C$ denote the set of sites within distance $l$ of $C$.
We will construct a sequence of operators $W_C^{(0)},W_C^{(1)},W_C^{(2)},\ldots,W_C^{(L)}$,
such that the last operator in the sequence is equal to $U^\dagger W_C U$ and such that each operator
$W_C^{(l)}$ is supported on $b_l(C)$.
Then, define $\kappa(W_C,l)=W_C^{(0)}$ for $l=0$ and $\kappa(W_C,l)=W_C^{(l)}-W_C^{(l-1)}$ for $l=1,2,\ldots$.
In this way, we decompose
\be
\label{decomp2}
U^\dagger W_C U = \sum_{l=0,1,\ldots,L} \kappa(W_C,l).
\ee
Then, we will bound $\Vert \kappa(W_C,l)\Vert$ by an exponentially decaying function of $l$.
Applying this decomposition (\ref{decomp2}) to each term in the decomposition of $W$ gives a decomposition of $W'$.
Once we have shown the exponential decay of $\Vert \kappa(W_C,l)\Vert$ as a function of $l$, the decay of $W'$ follows.

For any $j\in\Lambda$, any $a \in \{1,\ldots,n_{max}\}$, and any operator $W_C$, define ${\cal E}_{j,a}(W_C)=\frac{1}{2}(W_C+\gamma_{j,a} W_C \gamma_{j,a})$.  Let ${\cal E}_j(W_C)={\cal E}_{j,1}({\cal E}_{j,2}(\ldots {\cal E}_{j,n_{max}}(W_C) \ldots))$.
For any set of sites $S\subset \Lambda$, let
\be
{\cal E}_S(W_C)={\cal E}_{j_1}({\cal E}_{j_2}(\ldots {\cal E}_{j_{|S|}}(W_C)\ldots)),
\ee
where $j_1,\ldots,j_{|S|}$ are the sites in $S$ (the sequence of sites chosen does not matter).

For any bosonic operator $W_C$ and set $S$, note that ${\cal E}_S(W_C)$ is supported on the complement of $S$.
Set
\be
W_C^{(l)}={\cal E}_{\Lambda \setminus b_l(C)} (U^\dagger W_C U).
\ee
We will bound $\Vert W_C^{(l)} - U^\dagger W_C U\Vert$.  By a triangle inequality, this will give the desired bound on $\Vert \kappa(W_C,l) \Vert$.
We have
\begin{eqnarray}
\label{into}
\Vert U^\dagger W_C U - {\cal E}_S (U^\dagger W_C U))\Vert  & \leq &
 \sum_{j\in S} \sum_{a=1}^{n_{max}} \Vert [U^\dagger W_C U,\gamma_{j,a}] \Vert \\ \nonumber
&=& \sum_{j\in S} \sum_{a=1}^{n_{max}}  \Vert [W_C,U \gamma_{j,a} U^\dagger] \Vert.
\end{eqnarray}
We bound the commutator by
\begin{eqnarray}
\Vert [W_C,U\gamma_{j,a}U^\dagger] \Vert  &\leq & 
\Vert [W_C, \sum_{k\in C}\,  \sum_{b \in\{1,\ldots,n_{max}\}} O_{(j,a),(k,b)} \gamma_{k,b}] \Vert
\\ \nonumber
& \leq & 2 \Vert W_C \Vert \cdot \Vert \sum_{k\in C} \sum_{b \in\{1,\ldots,n_{max}\}} O_{(j,a),(k,b)} \gamma_{k,b}] \Vert \\ \nonumber
&\leq & 2 \Vert W_C \Vert \sqrt{\sum_{k\in C} (K_0 \exp(-\nu_0{\rm dist}(j,k)))^2}.
\end{eqnarray}
Substituting this into Eq.~(\ref{into}) and summing over $j,k,a$ gives the desired bound.
\end{proof}
\end{lemma}
We remark that the idea of using a map such as ${\cal E}_{\Lambda\setminus b_l(C)}$ to approximate an operator by some other operator supported on some given set $b_l(C)$ above is similar to an idea introduced in Ref.~\onlinecite{bhv} for the same purpose; here we use a sum while there an averaging over Haar unitaries was used.

\section{Stability of $\He$}
\label{stabHempty}
We now show theorem \ref{emptystable}.
Let us rescale the Hamiltonian by a constant so that $\Delta=1$.

Much of our proof repeats that of Ref.~\onlinecite{bh}; equation and lemma numbers here refer to the arXiv version.  The reader should be familiar with that paper to follow this section.

Let $H_{proj}$ be a Hamiltonian which is a sum of commuting projectors each supported on a cube $C$ of linear size $2$, so that $H_{proj}=\sum_C Q_{C}$, and let $P=\prod_{C} (1-Q_{C})$ be the projector onto the ground state subspace of $H_{proj}$.
Further, assume that $H_{proj}$ obeys conditions TQO-1,TQO-2 given in Ref.~\onlinecite{bh}.

Let the ground state of $H_{proj}$ be $g$-fold degenerate and let $\Delta$ be the difference between the $g$-th smallest and $(g+1)$-th smallest eigenvalues of $H_{proj}+V$.
The main theorem of \cite{bh} is
\begin{theorem}
\label{thm:mainBH}
There exist constants $J_0,c_1>0$ depending only on $\mu$
and the spatial dimension $D$ such that for all $J\le J_0$, if $V$ has $(J,\mu)$ decay, then  the spectral gap $\Delta$ of $H_{proj}+V$ is at least $1-c_1J-\delta$,
for some $\delta$ bounded by $J$ times a quantity decaying faster than any power of $L$.
\end{theorem}
In fact, Ref.~\onlinecite{bh} proves a stronger result, involving gaps between excited states; those results will not be relevant here. 
See also Ref.~\onlinecite{bhm}.

For the particular Hamiltonians $H_{proj}$ that we consider below, the constant $\delta$ will not be present, in fact.

We now show a generalization:
\begin{theorem}
\label{plusPos}
There exist constants $J_0,c_1$ depending only on $J,\mu,J_0,\mu_0,D$ such that
the following holds.
Let $H_0$ have $(J_0,\mu_0)$ decay and let $V$ have $(J,\mu)$ decay.  Assume $H_0 \geq H_{proj}$ for some $H_{proj}$ obeying the conditions above.  Assume $P H_0=0$.
Then, the spectral gap $\Delta$ of $H_0+V$ is at least $1-c_1J-\delta$,
for some $\delta$ bounded by $J$ times a quantity decaying faster than any power of $L$.
\end{theorem}
The inequality $H_0 \geq H_{proj}$ means that $H_0-H_{proj}$ is positive semi-definite.
Note that since $PH_0=0$, this means that $H_0$ and $H_{proj}$ have the same ground state subspace.

Theorem \ref{emptystable} then follows as a corollary by taking
$H_{proj}=\sum_{j \in \Lambda} n_j$ where $n_j=\psi^\dagger_j \psi_j$ and taking $H_0=\He$.  The assumption that $A$ has $[K,\nu]$ decay implies that
$H_0$ has $(J_0,\mu_0)$ decay for some $J_0,\mu_0$ depending only on $K,\nu$.
Note that $g=1$ in this case.

We now give some additional definitions that we need to prove theorem \ref{plusPos}.
\begin{definition}
A Hamiltonian $V$ is globally block-diagonal iff it preserves the ground subspace $P$, that is, $[V,P]=0$.
A Hamiltonian $V$ is locally block-diagonal iff all terms $V_{r,C}$ in the decomposition Eq.~(\ref{dec})
preserve the ground subspace, that is, $[V_{r,C},P]=0$ for all $r,C$.
\end{definition}

We now repeat some definitions verbatim (up to slight notational changes, in particular changing $A$ to $C$ everywhere since we use $A$ for something else here) from Ref.~\onlinecite{bh} that we will need in order to prove the results.
Consider a Hamiltonian
\be
\label{dec}
V=\sum_{r\ge 1} \; \; \sum_{C \in \calS(r)} \; \; V_{r,C},
\ee
where $V_{r,C}^\dag=V_{r,C}$ is an operator
acting non-trivially only on a cube $C$.\begin{definition}
A Hamiltonian $V$ has support near a site $u$ if all cubes in the decomposition Eq.~(\ref{dec})
contain $u$.
\end{definition}
\begin{definition}
A Hamiltonian $V$ has strength $J$ if there exists a function $f\, : \, \ZZ_+ \to [0,1]$ decaying faster than any power such that
\[
\| V_{r,C} \| \le J\, f(r) \quad \mbox{for all $r\ge 1$ for all $C\in \calS(r)$}.
\]
\end{definition}
Remark: of course, given any Hamiltonian $V$ on a finite size system, there is some function $f$ such that the above definition is obeyed.  However, the constants in the various lemmas of Ref.~\onlinecite{bh} depend upon the particular function.  Thus, the results are most useful when one considers a fixed function $f$ and then considers a family of Hamiltonians on increasing size all obeying this definition for some fixed function $f$, and then the bounds are uniform in system size.

Recall also that ${\cal D}_s$ will be defined to be a quasi-adiabatic evolution operator.  See Ref.~\onlinecite{lsmh} for quasi-adiabatic evolution and see Ref.~\onlinecite{bh} for the specific form of ${\cal D}_s$ that we use here. 

Finally, let us write $H_0=\sum_{u \in \Lambda} H_{0,u}$, where $H_{0,u}$ has support near $u$.
To obtain this decomposition, choose, for each cube $C$, some square $u(C)$ close to the center of $C$, and let $H_{0,u}$ be the sum, over all cubes whose center $u(C)=u$, of the term in the decomposition of $H_0$ supported on cube $C$.

We now prove theorem \ref{plusPos}.  Much of what follows is from Ref.~\onlinecite{bh} verbatim up to slight changes.
The reader should be familiar with that proof first.  The only important change is to the analysis of the difference $\tilde H_0-H_0$ later.

For any $s\in [0,1]$ define
\[
H_s=H_0+sV.
\]
Let $g$ be the ground state degeneracy of $H_0$. Let $E_{min}(s)$ be the smallest eigenvalue
of $H_s$ and let $E_{max}(s)$ be the $g$-th smallest eigenvalue of $H_s$ (taking into account multiplicities).
Finally, let $\Delta(s)$ be the spectral gap separating eigenvalues of $H_s$
in the interval $[E_{min}(s),E_{max}(s)]$ from the rest of the spectrum.
We shall choose the constants in the theorem sufficiently small such that
\be
\label{34bound}
\Delta(s) \ge  \frac34 \quad \mbox{for all $0\le s\le 1$}.
\ee
Suppose we have already proved the theorem under the additional assumption that
\be
\label{12bound}
\Delta(s)\ge \frac12 \quad \mbox{for all $s\in [0,1]$}.
\ee
We claim that this implies this would imply the theorem without this assumption:
in the remaining case there must exist $s^*\in [0,1)$ such that $\Delta(s)\ge 1/2$
for $s\in [0,s^*]$ and $\Delta(s^*) =1/2$ (use the fact that $\Delta(0)\ge 1$
and continuity of $\Delta(s)$).  Applying the theorem to a perturbation
$s^*V$ which satisfies Eq.~(\ref{12bound}) we conclude that $\Delta(s^*)\ge 3/4$
obtaining a contradiction.
Thus it suffices to prove the theorem for the case Eq.~(\ref{12bound}).

Define
\be
H_s'=U_s^\dag (H_0+sV) U_s=U_s^\dag H_s U_s,
\ee
where
\be
\label{Us}
U_s \equiv {\cal S}' \; \exp(i\int_0^s {\rm d}s' {\cal D}_{s'} ),
\ee
is the exact adiabatic continuation operator constructed in Section 6 of Ref.~\onlinecite{bh} and ${\cal S}'$ denotes an $s'$-ordered exponential.
Since we assumed $\Delta(s)\ge 1/2$ for all $s$, Lemma 6 of Ref.\onlinecite{bh}
implies that
\be
\label{com1}
[H_s',P]=0.
\ee
We can represent $H_s'$ as
\be
H_s'=H_0+V',
\ee
where
\be
V'=U_s^\dag H_0 U_s - H_0 + sU_s^\dag VU_s.
\ee
Lemma 2 of Ref.~\onlinecite{bh} implies that $\calD_s$ has strength $O(J)=O(1)$.
Applying Lemma 1 of Ref.~\onlinecite{bh} we conclude that $sU_s^\dag VU_s$
has strength $O(J)$.
Let us now focus on the term $U_s^\dag H_0 U_s - H_0$. We use
\be
U_s^\dagger H_0 U_s-H_0 = -i\int_0^s {\rm d}s'  \, U_{s'}^\dag [{\cal D}_{s'},H_0] U_{s'}.
\ee
Since $\calD_{s'}$ has strength $O(J)$,  the commutator $[{\cal D}_{s'},H_0]$
also has strength $O(J)$. Applying Lemma 1 of Ref.~\onlinecite{bh} to the unitary evolution
$U_{s'}^\dag$, we infer that $U_s^\dagger H_0 U_s-H_0$ has strength $O(J)$.
To conclude, we have shown that $V'$ has strength $O(J)$, and Eq.~(\ref{com1}) implies
\be
[V',P]=0,
\ee
that is, $V'$ is a globally block-diagonal perturbation with strength $O(J)$.
In Lemma~\ref{rewrite1} below, we will show that we can rewrite
\be
\label{rw1}
H_s'=H_0+V'=
H_0 + \sum_{u\in \Lambda} X_u
\ee
where $X_u$ obeys $\quad [X_u, P]=0$,
where $X_u$ has strength $O(J)$ and its support is near $u$.
Applying
Lemma 3 of Ref.~\onlinecite{bh} to Eq.~(\ref{rw1}) implies that $H_s'$ can be written in the form
\be
\label{lbddec}
H_s'=H_0+V'=H_0+V''+\Delta',
\ee
where $V''$ is locally block-diagonal with strength $O(J)$ and $\|\Delta'\|$
decays faster than any power of $L^*$ where the length $L^*$ is defined in Ref.~\onlinecite{bh}.  In fact, the term $\Delta'$ does not appear for the specific Hamiltonians considered in the present paper, so one may ignore this term
 (the notation $\Delta$ instead of $\Delta'$ was used in Ref.~\onlinecite{bh}; we use $\Delta'$ to avoid notational overload).

The statement of the theorem then follows from Lemma 5 of Ref.~\onlinecite{bh},
which implies that for such a locally block-diagonal perturbation $V''$, the eigenvalues of $H_{proj}+V''$ are contained in the set $\{0\} \cup [1-O(J),\infty)$ up to an overall energy shift.
Further,
the smallest eigenvalue of $H_{proj}+V''$ in the subspace orthogonal to the range of $P$ is lower bounded by $1-O(J)$, so that the zero eigenspace of $H_{proj}+V''$
coincides with the range of $P$ for sufficiently small $J$.
(This follows from the statement of Lemma 5 of Ref.~\onlinecite{bh} by a continuity argument increasing the strength of the perturbation $V''$ from $0$.)
 Then, since
$H_0\geq H_{proj}$ and hence $H_0+V'' \geq H_{proj}+V''$, the smallest eigenvalue of $H_{0}+V''$ in the subspace orthogonal to the range of $P$ is also lower bounded by $1-O(J)$.

The Lemma~\ref{rewrite1} that we need follows the idea in~\cite{kitaev2006anyons} to write a Hamiltonian of a gapped system as a sum of terms such that the ground states are
eigenvectors of each term separately (in \cite{solvloc} a related idea of writing it so that the ground state was an approximate eigenvector
of each term separately was considered).  The properties of $H_s'$ that we use are that it is globally block-diagonal, it has
a spectral gap $\geq 1/2$,
the perturbation $V'$ has strength $J$, and that it is unitarily related by $U_s$ to a Hamiltonian with $(J,\mu)$ decay.
\begin{lemma}
\label{rewrite1}
Let $H_s'$ be defined as above.  Then, we can re-write
\be
H_s'=H_0+ \sum_{u\in \Lambda } X_u
\ee
where $X_u$ obeys $\quad [X_u, P]=0$,
where $X_u$ has strength $O(J)$ and its support is near $u$.
\begin{proof}
We start from representing $V'$ as $V'=\sum_{u\in \Lambda} V_u$, where
$V_u$ includes only interactions affecting a site $u$. Then $V_u$ has strength $J$ and its support is near $u$.
We set
\be
\tilde{V}_u = \int_{-\infty}^\infty  dt \, g(t) \exp(i H_s' t) V_u \exp(-i H_s' t),
\ee
where $g(t)$ is a function satisfying $g(-t)=g(t)^*$  such that its Fourier transform $\tilde g(\omega)$ is infinitely differentiable, has $\tilde g(0)=1$, and
$\tilde g(\omega)=0$ for $|\omega|\geq 1/2$.
Define
\be
\tilde H_{0,u}= \int_{-\infty}^\infty  dt\, g(t)  \exp(i H_s' t) H_{0,u} \exp(-i H_s' t).
\ee
Then,
\be
H_s'=\int_{-\infty}^\infty dt\, g(t)  \exp(i H_s' t) H_s' \exp(-i H_s' t)=
\sum_{u \in \Lambda} \tilde H_{0,u} + \sum_{u\in \Lambda}  \tilde{V}_u.
\ee
By construction of $\tilde{g}(\omega)$ we have $(1-P) \tilde H_{0,u} P = (1-P) \tilde{V}_u P =0$.
Hence both $\tilde H_{0,u}$ and $\tilde{V}_u$ preserve $P$.
Using the definition of $\tilde{V}_u$ we get
\be
\label{tildeVu}
U_s \tilde{V}_u U_s^\dag  = \int_{-\infty}^\infty  dt \, g(t) \exp(i H_s t) U_s V_u U_s^\dag  \exp(-i H_s t).
\ee
Recall that $\calD_s$ has strength $O(1)$.
Thus we can apply Lemma 1 of Ref.~\onlinecite{bh}
to the unitary evolution $U_s$ to infer that $U_s V_u U_s^\dag$ has strength $O(J)$.
Because $\tilde g(\omega)$ is infinitely differentiable, $g(t)$ decays faster than any power.
Also, by assumptions of the theorem,  $H_s$ is a sum of terms with $(J_0,\mu_0)$ decay and $(J,\mu)$ decay.
Hence we can apply Lemma 2 of Ref.~\onlinecite{bh}
to the unitary evolution $\exp(i H_s t)$ to infer that $U_s \tilde{V}_u U_s^\dag$ has strength $O(J)$.
Finally, applying Lemma 1 of Ref.~\onlinecite{bh} to the unitary evolution $U_s^\dag$
we infer that $\tilde{V}_u$ has strength $O(J)$.
In addition, $\tilde{V}_u$ has support near $u$ since all Hamiltonians obtained at the intermediate steps
have support near $u$, see Lemmas 1,2 of Ref.~\onlinecite{bh}.

We now consider the terms $\tilde H_{0,u}$.  
Define
\be
M_u=\int_{-\infty}^\infty   dt\, g(t)  \exp(i H_0 t) H_{0,u} \exp(-i H_0 t).
\ee
This definition of $M_u$ differs from the definition of $\tilde H_{0,u}$ in that we use $H_0$ in the exponentials, rather than $H_s'$.
Note that $\sum_u M_u=H_0$.
We have
\begin{eqnarray}
\tilde H_{0,u} & = & \int_{-\infty}^{\infty} dt\,g(t) \exp(i H_s' t) H_{0,u} \exp(-i H_s' t)  \\ \nonumber
&=& M_u+ i\int_{-\infty}^{\infty} dt\, g(t) \int_0^t d t_1\,  \exp{(i H_s' t_1 )} [V',\exp(i H_0 (t-t_1)) H_{0,u} \exp(-i H_0 (t-t_1))] \exp{(-iH_s' t_1)}.
\end{eqnarray}
It follows that
\be
U_s (\tilde H_{0,u} -M_u)U_s^\dag = i\int_{-\infty}^{\infty} dt\, g(t) \int_0^t d t_1\,  \exp{(i H_s t_1 )} U_s [V',\exp(i H_0 (t-t_1)) H_{0,u} \exp(-i H_0 (t-t_1))] U_s^\dag \exp{(-iH_s t_1)}.
\ee
Let us just consider the case $t>0$ (the case $t<0$ is similar).  Then,
\begin{eqnarray}
&&\int_{0}^{\infty} dt\, g(t) \int_0^t d t_1\,  \exp{(i H_s t_1 )} U_s [V',\exp(i H_0 (t-t_1)) H_{0,u} \exp(-i H_0 (t-t_1))] U_s^\dag \exp{(-iH_s t_1)} \\ \nonumber
&=&\int_{0}^{\infty} dt_2\,  \int_0^\infty d t_1\, g(t_1+g_2) \, \exp{(i H_s t_1 )} U_s [V',\exp(i H_0 t_2) H_{0,u} \exp(-i H_0 t_2)] U_s^\dag \exp{(-iH_s t_1)}
\end{eqnarray}
Applying Lieb-Robinson bounds to the evolutions $\exp(i H_0 t_2),U_s,\exp(i H_s t_1),U_s^\dag$ successively, and using super-polynomial decay of $g(t_1+t_2)$,
we conclude that  $\tilde H_{0,u}- M_u$
has strength $O(J)$ and has support near $u$.  Remark: this bound slightly generalizes lemma 2 of Ref.~\onlinecite{bh}, as it involves integration over two different times, $t_1,t_2$; however, it is proven in the same way as that lemma.
In addition, $\tilde{H}_{0,u}- M_u$ commutes with $P$ since each term in the difference commutes with $P$ separately; note that 
$[M_u,P]=0$ since $P$ projects onto the ground state subspace of $H_0$ and $H_0$ has a spectral gap.

Let us define
$X_u=\tilde{V}_u + (\tilde H_{0,u}-M_u)$.
\end{proof}
\end{lemma}

\section{Remarks}
\label{remarks}
We have shown stability of free fermi Hamiltonians assuming a gap and exponential decay of interactions.  Let us make a few remarks.

First, as mentioned above, we assume either one or two Majorana operators per site.  One can handle the case of multiple Majorana operators per site by a rescaling of decay constants $\mu,\nu$ as mentioned.  However, as noted, if we simply assume that $A$ has $[K,\nu]$ decay and $V$ has $(J,\mu)$ decay, without assuming a bound on the number of Majorana operators per site, it seems that a Lieb-Robinson bound does not follow from known results.  The problem is that if we consider the terms in $H_{FF}$ restricted to some box, the bound on the operator norm of that operator is equal to the sum of positive eigenvalues of $A$ restricted to that box, and hence depends upon the number of modes.  We leave this problem open, but we expect that a bound can be proven, i.e., we would like a Lieb-Robinson bound which is uniform in the number of Majorana operators per site.
Note that all the various stability results such as \cite{bhm,bh,mff} all rely on Lieb-Robinson bounds so having such a bound is crucial if one desires uniformity with respect to the number of Majorana operators per site.

One application of theorem \ref{main} is to Hall conductance quantization.  In Ref.~\onlinecite{hmhall}, it is shown that Hall conductance is equal to an integer (up to almost exponentially small corrections) for gapped systems with local interactions.  Given the stability results here, this implies that if $H_{FF}$ has a given Hall conductance, the Hall conductance of $H_{FF}+V$ is the same for sufficiently small $J$ up to almost exponentially small corrections.  See also Ref.~\onlinecite{giuliani2016hall}.

Finally, let us remark on our choice to prove theorem \ref{plusPos} by modifying the proof of \cite{bh}.  This requires a small amount of work, but has the advantage that the perturbed $H_{doub}$ trivially satisfies the conditions of the theorem; we could instead have used existing results on frustration-free Hamiltonians but then we have to do a slight amount of additional work to write $H_{doub}$ as a sum of local frustration-free terms.  Note that a general Hamiltonian $H_0$ satisfying the conditions of theorem \ref{plusPos} is not necessarily given as a sum of frustration-free terms.  For example, $\He$ may have some terms $\psi^\dagger_j \psi_k$ for a far-separated pair of sites $j,k$ (with an exponentially small coefficient in this term); this term vanishes on the empty state, but it is minimized by a state with a single particle resonating between sites $j,k$.  One can adapt frustration-free results by a more elaborate procedure: first, drop terms beyond a certain range (by moving them to $V$ and treating them as a perturbation) and handle short-range terms by a ``windowing procedure".  Let us briefly sketch this windowing procedure, omitting all details.  Break the system up into a overlapping cubes (near the boundary of each cube, it overlaps with other cubes).  Write $\He=H_{diag}+H_{od}$, where $H_{diag}=\sum_{j \in \Lambda} \psi^\dagger_j \psi_j$.  Then, we define a free fermi Hamiltonian on each cube so that the sum of these Hamiltonians is equal to $H_{FF}$.  In each cube, we make the terms in $H_{od}$ decay rapidly near the boundary (so, near the boundary of one cube $C$, the terms in $H_{od}$ go to zero and the boundary of that cube is in the interior of some other cube $C'$, and the terms in $H_{od}$ supported near there will all be in $C'$.)  A term $\psi^\dagger_j \psi_j$ will be divided with equal weight among all cubes containing site $j$.  By doing this, we make it so that on each cube, the Hamiltonian in that cube is minimized by the empty state, giving a frustration-free Hamiltonian.  As one can see, this method also requires some work, which is why we use the method here.

{\it Acknowledgments---} I thank S. Michalakis for very useful discussions.  I thank S. Bravyi for comments on a draft of this paper.
\bibliography{ffbib}
\end{document}